\documentclass[12pt,a4paper]{article}
\usepackage{amsmath,amssymb,mathrsfs,framed,esint,slashed,braket}
\usepackage[colorlinks]{hyperref}
\usepackage{color}
\usepackage[all]{xy}
\usepackage{graphicx}

\newcommand{\rem}[1]{}

\newcommand{\de}{{\rm d}}

\newcommand{\bq}{{\boldsymbol{q}}}

\newcommand{\bp}{{\boldsymbol{p}}}

\newcommand{\tA}{{\,\widetilde{\!\boldsymbol{\cal A}\,}}}

\newcommand{\bx}{{\mathbf{x}}}

\newcommand{\br}{{\boldsymbol{r}}}

\newcommand{\bA}{{\boldsymbol{\cal A}}}

\newcommand{\bu}{{\boldsymbol{u}}}
\newcommand{\bxi}{{\boldsymbol{\xi}}}

\newcommand{\ben}{\begin{eqnarray}}
\newcommand{\een}{\end{eqnarray}}

\begin{document}

\title{
Regularized Born-Oppenheimer molecular dynamics
}
\author{Jonathan I. Rawlinson$^1$, Cesare Tronci$^{2,3}$
\smallskip
\\ 
\small
$^1$\it School of Mathematics, University of Bristol, Bristol, UK
\\
\small
$^2$\it Department of Mathematics, University of Surrey, Guildford, UK
\\
\small
$^3$\it Department of Physics and Engineering Physics, Tulane University, New Orleans LA, USA}
\date{}
\maketitle

\begin{abstract} 
While the treatment of conical intersections in molecular dynamics generally requires nonadiabatic approaches,  the Born-Oppenheimer adiabatic approximation is still  adopted as a valid alternative in certain circumstances. In the context of Mead-Truhlar minimal coupling, this paper presents a
 new closure of the nuclear Born-Oppenheimer equation, thereby leading to a molecular dynamics scheme capturing geometric phase effects. Specifically,  a semiclassical closure of the nuclear Ehrenfest dynamics is obtained through a convenient prescription for the nuclear Bohmian trajectories.  The conical intersections are suitably regularized in the resulting nuclear particle motion and the associated Lorentz force involves a smoothened Berry curvature identifying a loop-dependent geometric phase. In turn, this geometric phase rapidly reaches the usual topological index as the loop expands away from the original singularity. This feature reproduces the phenomenology appearing in recent exact nonadiabatic studies, as shown explicitly in the Jahn-Teller problem for linear vibronic coupling. Likewise, a newly proposed regularization of the diagonal correction term is also shown to reproduce quite faithfully the energy surface presented in  recent nonadiabatic studies.

\end{abstract}

{\small
\tableofcontents
}

\bigskip

\section{Introduction}

Following the long-standing success of Born-Oppenheimer molecular dynamics over several decades, more recently a great deal of work has gone into the design of nonadiabatic molecular dynamics schemes benefiting from increasing computational power. The development of new simulation codes for nonadiabatic dynamics is mostly motivated by the fact that the usual adiabatic Born-Oppenheimer factorization of the molecular wavefunction is generally known to break down at conical intersections, where nonadiabatic transitions may indeed occur. Nevertheless, the adiabatic theory remains in principle a valid option whenever the energy of the conical intersection is sufficiently higher than the energy associated to nuclear motion, so that nonadiabatic transitions become unlikely. However, the presence of conical intersections makes the adiabatic theory extremely challenging due to the singularities appearing in the nuclear Schr\"odinger equation. Indeed, the standard quantum-classical picture from Born-Oppenheimer molecular dynamics leads to intractable trajectory equations involving $\delta$-like Lorentz forces. Then, the study of molecules with a large number of atoms still represents a major challenge, which stands as the motivation of this paper.


\subsection{Born-Oppenheimer molecular dynamics}\label{Sec:BornOppenheimer}

As explained in, for example, \cite{EntropyReviewNonadiabatic,Marx}, the  molecular wavefunction $\Psi(\{\br\},\{\bx\},t)$ for a system composed of $N$ nuclei with coordinates $\br_i$ and $n$ electrons with coordinates $\bx_a$ is factorized in terms of a nuclear wavefunction $\Omega(\{\br\},t)$ and a time-independent electronic function $\phi(\{\bx\};\{\br\})$ depending parametrically on the nuclear coordinates $\{\br\}_{i=1\dots N}$. For the sake of simplicity, here we shall consider the simple case of a single electron and nucleus, so that
$\Psi(\br,\bx,t)= \Omega(\br,t)\phi(\bx;\br)$ and $\int|\phi(\bx;\br)|^2\,\de^3x=1$, where the latter is enforced by the normalizations of $\Psi$ and $\Omega$. Equivalently, upon making use of Dirac's notation, we denote $\ket{\phi(\br)}:=\phi(\bx;\br)$ and write
\begin{align}
  \Psi(t) &= \Omega(\br,t)\ket{\phi(\br)}\label{StandardBOAnsatz}.
\end{align} 
The partial normalization condition becomes  $\|\phi(\br)\|^2:=\braket{\phi(\br)|\phi(\br)}=1$ and the Hamiltonian operator for the system reads $\widehat{H}=-\hbar^2M^{-1} \Delta/2\ +\widehat{H}_e$. Here,  $M$ is the nuclear mass and all derivatives are over the nuclear coordinate $\br$. In addition,  the electronic state is taken as the fundamental eigenstate of the electronic Hamiltonian $\widehat{H}_e$, so that $\widehat{H}_e\ket{\phi(\br)}=E(\br)\ket{\phi(\br)}$ and the eigenvalue identifies the {\it adiabatic energy surface}. 
The motivation for such an ansatz comes from the separation of molecular motion into fast and slow dynamics due to the large mass difference between that of the electron and nucleus \cite{BornOppenheimer1927,Hagedorn}. 
 On the other hand, the nuclear wavefunction obeys the Born-Oppenheimer equation
\begin{equation*}
i\hbar\frac{\partial\Omega}{\partial t}=\left[\frac{(-i\hbar\nabla+ \boldsymbol{\cal A})^2}{2M}+\varepsilon(\phi,\nabla\phi)\right]\Omega
\,,
\end{equation*}
where we have introduced the {\it Berry connection} \cite{Berry1984}
$\boldsymbol{\cal A}(\br):=\braket{\phi|-i\hbar\nabla\phi}$
and   the 
{\it effective electronic potential}
\begin{align}\nonumber
  \varepsilon(\phi,\nabla\phi):=&\ E+\frac{\hbar^2}{2M}\|\nabla\phi\|^2 
  - \frac{|\boldsymbol{\cal A}|^2}{2M}
  \\
  =&\ E+\frac{\hbar^2}{4M}\|\nabla\rho_\phi\|^2
  \label{effectivepotential}\,.
\end{align} 
In the second line, we have introduced the  parameterized density operator $\rho_\phi(\br)=|\phi(\br)\rangle\langle\phi(\br)|$. 
Then, upon defining $\boldsymbol{\cal B}=\nabla\times\boldsymbol{\cal A}$, the classical limit obtained via the usual Hamilton-Jacobi analogy leads to the trajectory equation
\[
M\,\ddot{\!\bq}+{\,\dot{\!\bq}}\times\boldsymbol{\cal B}=-\nabla \varepsilon(\phi,\nabla\phi)
\,,
\]
which is further simplified by neglecting the so-called \emph{diagonal correction} $\hbar^2M^{-1}\|\nabla\phi\|^2/2= -\hbar^2M^{-1}\text{Re}\braket{\phi|\Delta\phi}/2$, see \cite{TullyNonadiabaticDynamics}.
In addition, since $|\phi\rangle$ is typically real-valued we have $\boldsymbol{\cal A}(\br)\equiv0$, thereby recovering the usual quantum-classical picture of classical nuclei evolving on the electron energy surface $E(\br)$ emerging from the quantum electronic problem. {Then, while the electrons retain their quantum wavefunction description, nuclear motion is described in terms of finite-dimensional classical trajectories evolving on a single adiabatic potential energy surface.}

 While the fundamental nature of quantum-classical coupling represents one of the most challenging open problems in physics and chemistry \cite{AgCi07,CrBa18,BoGBTr19,GBTr20}, Born-Oppenheimer molecular dynamics has been extremely successful in a variety of contexts \cite{Marx} and remains the most established theory in quantum chemistry.

\subsection{Generalized Born-Oppenheimer theory}

The presence of singularities (\emph{conical intersections}) in the electronic potential energy surface (PES) leads to double-valued electronic wavefunctions {thereby  posing several challenges}. The name `conical intersection' is due to the fact that separate eigenvalues in the electronic spectral problem may intersect for specific nuclear coordinates, which emerged historically as points where the energy surfaces form the shape of a double cone \cite{Teller37,Yarkony}. In 1979, Mead and Truhlar \cite{MeadTruhlar1979} exploited gauge transformations in order to make electronic wavefunctions again single-valued and restore the Born-Oppenheimer separation \eqref{StandardBOAnsatz} of nuclear and electronic dynamics. 

Upon exploiting the invariance of the Born-Oppenheimer factorization \eqref{StandardBOAnsatz} under the gauge transformation 
\begin{equation}\label{GInv}
\Omega\mapsto e^{-i\zeta/\hbar}\Omega
\,,\qquad\qquad\quad
|\phi\rangle\mapsto 
e^{i\zeta/\hbar}|\phi\rangle
\,,
\end{equation}
the Mead-Truhlar {minimal coupling} method selects the phase function $\zeta(\br)$ in such a way that  the new electronic state $|\phi'\rangle=e^{i\zeta/\hbar}|\phi\rangle$ is 
single-valued \cite{Kendrick2003,Mead1992} and so avoids the need to deal with double-valued functions. However, since $|\phi\rangle$ is real and the phase  $\zeta$  has a pole at the conical intersection, this approach leads to the introduction of a highly singular magnetic potential $\boldsymbol{\cal A}' = \nabla\zeta$  \cite{Mead1992,Kendrick2003,Kuppermann}.
Then, the nuclear motion obeys the \emph{generalized Born-Oppenheimer equation} \cite{Kendrick2003,Mead1992}
\begin{equation}\label{genBOeq}
i\hbar\frac{\partial\Omega}{\partial t}=\left[\frac{(-i\hbar\nabla+ \nabla\zeta)^2}{2M}+\varepsilon(\phi',\nabla\phi')
\right]\Omega
\,.
\end{equation}
Here, we have dropped the prime on $\Omega'=e^{-i\zeta/\hbar}\Omega$ to avoid proliferation of notation.
Notice that neglecting the diagonal correction leads to {replacing the effective potential $\varepsilon(\phi',\nabla\phi')$ by the energy surface $E$}. Indeed, we have $\varepsilon(\phi',\nabla\phi')=\varepsilon(\phi,\nabla\phi)= E+\hbar^2M^{-1}\|\nabla\phi\|^2/2$ since  \eqref{effectivepotential} is gauge-invariant and $\phi$ is real.
{
Upon neglecting the diagonal correction, equation \eqref{genBOeq} has been considered as a simplifying alternative to nonadiabatic approaches retaining quantum electronic transitions \cite{Kendrick2018}. The latter become more and more likely as the  energy associated to nuclear motion approaches the energy  of the conical intersection. Otherwise, the adiabatic assumption remains valid and recently Kendrick and collaborators used equation \eqref{genBOeq} to emphasize the role of the geometric phase in ultracold chemical reactions \cite{KeHaBa15}. In addition, new approaches have recently been proposed in \cite{MaZhGuYa16} to construct the phase function $\zeta$ in general higher-dimensional configurations and this methodology was \makebox{applied to situations involving polyatomic systems \cite{XiMaYaGu17}.}
} 

Importantly, since both the vector potential $\nabla\zeta$ and the electronic potential are now singular, the nuclear problem is essentially equivalent to the Aharonov-Bohm problem, thereby leading to a particularly relevant Berry phase effect, known as the  \emph{molecular Aharonov-Bohm effect}.
In this case, $\nabla\times\nabla\zeta$ is usually represented as a delta function \cite{Kendrick2003,Kleinert} and thus the standard Hamilton-Jacobi analogy leads to intractable classical trajectory equations which prevent a molecular dynamics approach and lead to the necessity of solving the Schr\"odinger equation \eqref{genBOeq} for the entire nuclear wavefunction \cite{Kendrick2003,Wyatt}. In this context, a standard approach is to approximate and truncate an expansion of $\Omega$ over the basis set provided by Gaussian coherent states \cite{Heller,Littlejohn,Perelomov}. If ${\cal G}$ is a normalized Gaussian and $\chi(\br;\bq,\bp)=\sqrt{{\cal G}(\br-\bq)}\exp\!\big[i\hbar^{-1}\bp\cdot(\br-\bq/2)\big]$ is the corresponding coherent state parameterized by $(\bq,\bp)$, then the nuclear wavefunction can be expressed as
\[
\Omega(\br,t)=\int\! c(\bq,\bp,t)\chi(\br;\bq,\bp)\,\de^3q\,\de^3p\simeq \sum_{k=1}^Nc_k(t) \chi(\br;\bq_k(t),\bp_k(t))
\,.
\]
The second step involves an approximation and subsequent truncation of the coherent state basis \cite{HuHe88}. Despite its accompanying issues \cite{ShCh04}, this type of representation is the basis of most current models in molecular dynamics \cite{JoIz18}. However, this representation leads to cumbersome equations of motion, which are often simplified by specifically devised methods {and uncontrolled approximations} \cite{JoSiRyIz16}. Eventually, this equivalent fully quantum nuclear dynamics still involves major challenges which arise from the singular character of $\zeta$ and $E$ and this represents an important problem especially in the study of molecules with a large number of nuclei. On the other hand, an exact dynamical study \cite{ScEn19} has recently shown that, at least in the case of the Shin-Metiu model \cite{ShMe95}, classical trajectories adequately reproduce  the quantum nuclear dynamics within the Born-Oppenheimer approximation. Then, following the long-standing success of Born-Oppenheimer molecular dynamics, this result  provides further motivation for the development of an adiabatic quantum-classical model retaining geometric phase effects.

Due to the singular character of $\zeta$,  the nuclear Berry phase
\[
\Gamma=\oint_{\gamma_0\!} \nabla\zeta(\br)\cdot\de\br
\]
 has an intrinsic topological nature and it is actually expressed as a topological index. While this topological character is well established in the standard Aharonov-Bohm effect \cite{Aharonov}, its manifestation {has recently been debated in \cite{Gross1,Gross2,Gross3}. Therein,} Gross and collaborators have shown that the molecular Berry phase  may be considered as a manifestation of a geometric (path-dependent) effect. According to these studies, the value of the geometric phase rapidly tends to the  {topological index} as the loop encircling the conical intersection becomes more and more distant from the singularity. These studies involve the exact factorization of the molecular wavefunction, where the electronic factor in \eqref{StandardBOAnsatz} is allowed to be time-dependent. Gross and collaborators interpret this geometric deviation of the Berry phase from the topological index value as a \emph{non-adiabatic effect}, since the exact factorization framework {deals with non-adiabatic dynamics}.
 
 \subsection{Goal of the paper}
 
This paper proposes a new approach to the formulation of adiabatic molecular dynamics in the presence of conical intersections. While the generalized Born-Oppenheimer theory essentially prevents the possibility of treating nuclei as classical particles, here we shall recover the usual trajectory ensemble picture from  quantum-classical molecular dynamics by resorting  to Bohmian trajectories. Without attempting to solve the quantum nuclear problem and instead of adopting  the classical limit, we shall obtain a molecular dynamics scheme by  performing a semiclassical closure involving the nuclear Bohmian trajectories. Motivated by the Ehrenfest theorem for quantum nuclear dynamics, this closure scheme exploits the hydrodynamic analogy between Bohmian trajectories and Lagrangian fluid paths,  which are restricted to define translations in the nuclear coordinate space.

As we shall see, this intermediate quantum-semiclassical picture shows that the singular character of the electronic potentials becomes manifest in the nuclear evolution  only when setting $\hbar=0$, which then leads to intractable classical trajectory equations. On the other hand, a nonzero value of Planck's constant (no matter how small) leads to regular {nuclear} trajectories evolving under the influence of smoothened effective potentials given by the convolution of the original potentials with the nuclear probability density. In turn, the removal of the singularities makes the Berry phase a geometric (path-dependent) quantity in analogy with  the recent findings in \cite{Gross2,Gross3}. As we shall see in the simple case of Jahn-Teller systems, the geometric phase resulting from the Bohmian closure successfully reproduces the phenomenology appearing in exact nonadiabatic studies \cite{Gross3}: starting from zero, the Berry phase rapidly tends to the Aharonov-Bohm topological index as the loop encircling the conical intersection becomes distant from the singularity.


A further step taken in this paper concerns the diagonal Born-Oppenheimer correction term. As is well known,  unlike the singularity in the Berry connection, the  singularity in the diagonal correction term is not integrable and leads to a divergent total energy unless the nuclear density vanishes at the conical intersection. As discussed in \cite{MeLe16}, this extremely singular behavior is absent in exact nonadiabatic treatments \cite{AbediEtAl2012} and thus is an artifact of the Born-Oppenheimer approximation.  While exact treatments show that the diagonal correction  is negligible in many situations, this is not always the case \cite{MeLe16} and occasionally one is led to formidable challenges in dealing with the singular character of this term. Here, following the methodology introduced in this paper, we shall develop a new gauge-invariant regularization approach for dealing with the diagonal correction. Specifically, we shall perform a regularization of the density matrix form of the electronic potential \eqref{effectivepotential} and we shall see how this operation allows one to treat the diagonal correction without having to face divergent integrals.

\rem{ 
While the above approaches associate a single trajectory to each nucleus, we shall present a new approach exploiting coupled multiple trajectories to provide a dynamical sampling of the nuclear ensemble. Based on previous work by the authors \cite{FoHoTr19}, a density matrix formulation will be employed to formulate the classical limit in the hydrodynamic treatment of mixed states. Then, as we shall show, combining the hydrodynamic Clesch representation  \cite{} with quantum coherent states leads to an ensemble of coupled trajectory equations that can be used to sample nuclear ensemble dynamics in the presence of conical intersections. The relation of this approach to the previously introduced \emph{Bohmion method} \cite{FoHoTr19} will also be presented explicitly.
} 

\section{From expectation values to Bohmian trajectories}
As we discussed above, the presence of singularities on the electronic energy surface can ruin the usual quantum-classical picture of molecular dynamics. This is due to the emergence of a delta-like Berry curvature rendering the classical trajectory equations intractable. In the absence of singularities, the classical picture is  obtained by applying the usual WKB argument leading to the Hamilton-Jacobi analogy. While this is probably the best-established way of obtaining the classical limit, here we shall consider an alternative approach based on Bohmian trajectories and inspired by the Ehrenfest dynamics of nuclear expectation values.

\subsection{Remarks on Ehrenfest nuclear dynamics\label{sec:Erehnf}}
As outlined in standard textbooks, the equations of motion for the expectation values $(\langle\widehat{\boldsymbol{Q}}\rangle,\langle\widehat{\boldsymbol{P}}\rangle)$ of the canonical observables $(\widehat{\boldsymbol{Q}},\widehat{\boldsymbol{P}})$ coincide with the classical trajectory equations under the assumption that $\langle A(\widehat{\boldsymbol{Q}},\widehat{\boldsymbol{P}})\rangle\simeq A(\langle\widehat{\boldsymbol{Q}}\rangle,\langle\widehat{\boldsymbol{P}}\rangle)$ for any observable $A$. 
The Ehrenfest approach to the classical limit is performed in two stages: 1) compute the Ehrenfest equations for the canonical observables, 2)  replace averages of functions by functions of averages (semiclassical assumption). While this approach hinges on interpretative arguments previously questioned in \cite{Ballentine}, it leads to exactly the same classical equations as the Hamilton-Jacobi analogy. In turn, upon denoting $\boldsymbol{\cal B}=\nabla\times\nabla\zeta$, this approach shows that the Ehrenfest equations associated to the quantum nuclear motion \eqref{genBOeq}
\[
M\frac{\de}{\de t}\langle\widehat{\boldsymbol{Q}}\rangle=\langle\widehat{\boldsymbol{P}}+\nabla\zeta\rangle
\,,\qquad\quad
\frac{\de}{\de t}\langle\widehat{\boldsymbol{P}}+\nabla\zeta\rangle+\frac1{2M}\big\langle(\widehat{\boldsymbol{P}}+\nabla\zeta)\times\boldsymbol{\cal B}-\boldsymbol{\cal B}\times(\widehat{\boldsymbol{P}}+\nabla\zeta)\big\rangle=
-\langle\nabla \varepsilon\rangle
\]
do not exhibit delta-like singularities, which instead are smoothened by the averaging process. Then, instead of taking the classical limit, these Ehrenfest equations may be simplified by assuming $\langle (\widehat{\boldsymbol{P}}+\nabla\zeta)\times\boldsymbol{\cal B}\rangle\simeq \langle \widehat{\boldsymbol{P}}+\nabla\zeta\rangle\times\langle\boldsymbol{\cal B}\rangle\simeq -\langle\boldsymbol{\cal B}\times(\widehat{\boldsymbol{P}}+\nabla\zeta)\rangle$. This assumption leads naturally to the following  equation for $\bq=\langle\widehat{\boldsymbol{Q}}\rangle$:
\begin{equation}\label{unclosedeqn}
M\ddot\bq+\dot\bq\times\langle\boldsymbol{\cal B}\rangle=-\langle\nabla \varepsilon\rangle
\,.
\end{equation}
However, the equation above still needs to be closed in such a way to express both $\langle\boldsymbol{\cal B}\rangle$ and $\langle\nabla \varepsilon\rangle$ as functions of $\bq$. Ordinarily, we would replace  $\langle\boldsymbol{\cal B}\rangle=\boldsymbol{\cal B}(\bq)$ and $\langle\nabla \varepsilon\rangle=\nabla \varepsilon(\bq)$ to recover the classical limit. However, this leads to intractable equations in the presence of singularities.
Thus, we are motivated to look for an alternative closure that preferably reduces to  classical dynamics in some limit case. 

\subsection{Bohmian approach to nuclear trajectories\label{sec:CS}}

In this section, a suitable closure of equation \eqref{unclosedeqn} is provided by operating within the hydrodynamic Bohmian picture. This method is  inspired by previous work in plasma physics \cite{Ho83} and geophysical fluid dynamics \cite{Ho86} and its application to Born-Oppenheimer dynamics was recently outlined  in \cite{FoTr20}. Also, this procedure {emerges as the Bohmian counterpart to} the coherent state approach used in \cite{KramerSaraceno,Shi,ShiRabitz}. Therein, the Dirac-Frenkel variational principle underlying the Schr\"odinger equation is modified by restricting the wavefunction to evolve as a Gaussian coherent state \cite{Littlejohn, Perelomov}; see \cite{BLTr15,BLTr16} for a geometric characterization of this approach. Here, instead of prescribing  the nuclear wavefunction evolution {(which would break gauge invariance)}, we shall adopt an alternative closure strategy by focusing on the hydrodynamic variational principle underlying nuclear Bohmian trajectories. More specifically, we shall modify the hydrodynamic variational principle  by restricting Bohmian trajectories to evolve according to translations in the nuclear coordinate space. As we shall see, this method amounts to prescribing a closure for the Ehrenfest dynamics of the expectation value of the nuclear position, thereby providing a closure to equation \eqref{unclosedeqn}.


We begin by considering the generalized Born-Oppenheimer equation \eqref{genBOeq} obtained by applying the Mead-Truhlar method. 
In the Bohmian approach, we write $\Omega(\br,t)=\sqrt{D(\br,t)}e^{iS(\br,t)/\hbar}$ and  define $M\bu=\nabla S+\nabla \zeta$, thereby leading to the following hydrodynamic Lagrangian for nuclear motion:
\begin{align}
\label{BO-Lagr3}
L(\bu,D)   &= \int D\left(\frac{M}{2}|\bu|^2 - \bu\cdot\nabla \zeta+ \frac{\hbar^2}{8M}\frac{|\nabla D|^2}{D^2} -
\varepsilon (\phi',\nabla\phi')\right)\,\text{d}^3r\,.
\end{align}
At this point, the fundamental variables become the the probability density $D$ and the hydrodynamic velocity $\bu$. In turn, the latter produces the Bohmian trajectories ${\boldsymbol\eta}(\br,t)$ in terms of Lagrangian fluid paths as
\[
\dot{\boldsymbol\eta}(\br,t)=\bu(\boldsymbol{s},t)\big|_{\boldsymbol{s}={\boldsymbol\eta}(\br,t)}
\,.
\]
In Bohmian mechanics, the evolution of these Lagrangian paths replaces the Schr\"odinger equation for the quantum wavefunction. In addition, Bohmian trajectories govern the dynamics of the probability density via the Lagrange-to-Euler map
$
D(\br,t)=\int \!D_0(\boldsymbol{s})\,\delta(\br-\boldsymbol\eta(\boldsymbol{s},t))\,\de^3r
$.
Notice that this construction produces the Euler-Poincar\'e variations \cite{HoMaRa98}
\[
\delta\bu=\dot\bxi+(\bxi\cdot\nabla)\bu-(\bu\cdot\nabla)\bxi
\,,\qquad\qquad
\delta D=-\operatorname{div}(D\bxi)
\,,
\]
where $\bxi$ is an arbitrary displacement vector field such that $\delta {\boldsymbol\eta}(\br,t)=\bxi(\boldsymbol{s},t)|_{\boldsymbol{s}={\boldsymbol\eta}(\br,t)}$.
\rem{ 
Now, denote $F=\int D_0(\br-\bq)\varepsilon (\tilde\phi,\nabla\tilde\phi)\,\de^3r$. Since 
\[
\delta{\tilde\phi}=-i\hbar^{-1}(\delta\Theta)\tilde\phi
\,,
\qquad\qquad
\delta\tA=\nabla\delta\Theta
\,,
\]
we have
\begin{align*}
{\delta F}=&\ \delta\bq\cdot\nabla F+\int\!\frac{\delta F}{\delta \Theta}\delta\Theta\,\de^3r
\\
=&\ 
\delta\bq\cdot\nabla F+\operatorname{Re}\int\!\frac{\delta F}{\delta \tilde\phi}^\dagger\delta\Theta(-i\hbar^{-1}\tilde\phi)\,\de^3r
\\
=&\ 
\delta\bq\cdot\nabla F+\hbar^{-1}\int\!\delta\Theta\operatorname{Im}\!\left\langle\frac{\delta F}{\delta \tilde\phi}\bigg|\,\tilde\phi\right\rangle\de^3r
\\
=&\ 
\delta\bq\cdot\nabla F
\,,
\end{align*}
where we have verified that $\operatorname{Im}\langle{\delta F}/{\delta \tilde\phi}\,|\,\tilde\phi\rangle\equiv0$.
Then, we have the Euler-Lagrange equation for $\Theta$:
\[
0=\partial_t D+\operatorname{div}\frac{\delta}{\delta \tA}\int\!D\bu\cdot\tA\,\de^3r=\partial_t D+\operatorname{div}(D\bu)
\]
} 
Instead of going ahead writing down the hydrodynamic equations, here we proceed 
by  formulating a finite-dimensional nuclear motion that can serve as a closure for  equation \eqref{unclosedeqn}. To this end,   we prescribe a specific form of Bohmian trajectory, that is
\begin{equation}\label{Bohmtraj}
\boldsymbol\eta(\br,t)=\br+\bq(t)
\,.
\end{equation}
This step amounts to restricting the infinite-dimensional  Bohmian trajectory $\boldsymbol\eta(\br,t)$ to evolve as a finite-dimensional translation in the nuclear coordinate space. { Notice that here other options are also available; for example, one could write $\boldsymbol\eta(\br,t)=M(t)\br+\bq(t)$ to allow squeezing. However, in this paper we restrict to translations only. Then, equation \eqref{Bohmtraj}} leads to
\[
\bu(\br,t)=\dot\bq
\,,\qquad\qquad\ 
D(\br,t)=D_0(\br-\bq(t))
\,,
\]
so that $\bq=\langle\widehat{\boldsymbol{Q}}\rangle$. Here, $D_0$ is the initial nuclear density and would normally be chosen as a Gaussian, although here one may allow for alternative choices such as generalized normal distributions of the type $D_0(\boldsymbol{x})\propto e^{\,|\boldsymbol{x}|^\alpha/\lambda}$. In general, the closure \eqref{Bohmtraj} takes the  Lagrangian \eqref{BO-Lagr3} into the form
\begin{align}\nonumber
  L(\bq,\dot{\bq})
 =&\  \frac{M}{2}|\dot{\bq}|^2 
   - {\bar\varepsilon (\phi',\nabla\phi')}-\dot\bq\cdot\!\overline{\nabla\zeta}(\bq)
   \,,
   \label{BO-Lagr4}
  \end{align}
where we have introduced the bar notation so that
\begin{align*}
\overline{\nabla\zeta}(\bq)=&\ \int \!D_0(\br-\bq){\nabla\zeta}(\br)\,\de^3 r
\,,\qquad\qquad\ 
{\bar\varepsilon (\phi',\nabla\phi')}=\int \!D_0(\br-\bq)\varepsilon (\phi',\nabla\phi')\,\de^3r
\,.
\end{align*}
Then, the nuclear equation of motion is as follows:
\begin{equation}
M\ddot\bq+\dot\bq\times\nabla\times\overline{\nabla\zeta}(\bq)=-\nabla{\bar\varepsilon (\phi',\nabla\phi')}
\,.
\label{Bohmclos}
\end{equation}
We notice that dropping the bars everywhere in \eqref{Bohmclos} would correspond to the classical limit, so that the closure proposed here can be understood as a type of semiclassical closure of equation \eqref{unclosedeqn}. As already commented above, the classical limit yields intractable equations in the presence of singularities since $\nabla\times\nabla\zeta$ is a delta-like magnetic field. Here, we  notice that  neglecting the diagonal correction leads to {replacing  $\bar\varepsilon(\phi',\nabla\phi')$ by the regularized energy surface $\,\overline{\!E}$} in \eqref{Bohmclos}, since $\bar\varepsilon(\phi',\nabla\phi')$ is gauge-invariant and $\phi$ is real. In this case,  we observe that  the semiclassical trajectory equation \eqref{Bohmclos} avoids the singularities, which are instead smoothened out by the probability density $D_0$ acting as a convolution kernel.  

It is perhaps useful to compare the trajectory equation \eqref{Bohmclos} with the equation of motion for the Bohmian trajectory \eqref{Bohmtraj}. Upon using the density transport equation $\partial_t D+\operatorname{div}(D\bu)=0$, these Bohmian trajectories obey the Euler-Lagrange equations
\begin{multline}\label{ELBohm}
D_0(\br)\dot{\boldsymbol\eta}(\br,t)+D_0(\br)\dot{\boldsymbol\eta}(\br,t)\times\nabla_{{\!\boldsymbol\eta}}\times\nabla\zeta(\boldsymbol{s},t)|_{\boldsymbol{s}={\boldsymbol\eta}}
\\
=-D_0(\br)\nabla_{\!\boldsymbol\eta}
\varepsilon (\boldsymbol{s},t)|_{\boldsymbol{s}={\boldsymbol\eta}}
- \frac{\hbar^2}{8M}\frac{\delta}{\delta{\boldsymbol\eta}}\int\frac{|\nabla D(\boldsymbol{s},t)|^2}{D(\boldsymbol{s},t)^2}\,\de^3s\,.
\end{multline}
Notice that the Lorentz force experienced by the Bohmian trajectory is produced by the magnetic field $\nabla_{{\!\boldsymbol\eta}}\times\nabla\zeta|_{\boldsymbol{s}={\boldsymbol\eta}}$, which is singular in the presence of conical intersections. If we now  replace \eqref{Bohmtraj}, the quantum potential reduces to an irrelevant constant and integrating over the nuclear position leads to \eqref{Bohmclos}. Thus, we observe that, while the Bohmian trajectory is subject to a delta-like Lorentz force, the latter appears smoothened in the Ehrenfest equation for the expectation value $\bq=\langle\widehat{\boldsymbol{Q}}\rangle$. The implications for the geometric phase are discussed in the following section.

\subsection{Geometric character of the Berry phase\label{sec:BP}}

In the mixed quantum-semiclassical picture associated to equation \eqref{Bohmclos}, the Berry phase effects occurring in the Bohmian trajectory evolution \eqref{ELBohm}   appear in Ehrenfest's nuclear trajectories as a Lorentz force producing a gyration around the static magnetic field $\nabla_{\!\bq\!}\times\overline{\nabla\zeta}(\bq)$. We notice how the presence of poles in $\zeta$ is absolutely necessary for this effect to appear; indeed, if $\Theta$ is a smooth function, a simple integration by parts yields $\overline{\nabla\Theta}=\nabla\,\overline{\!\Theta\!}\,$,
so that $\nabla\times\nabla\,\overline{\!\Theta\!}\,=0$. On the contrary, in the presence of singularities, non-vanishing boundary terms emerge from the integration thereby contributing to the magnetic field. A very similar situation also occurs in the Ehrenfest equations associated to the Aharonov-Bohm problem; see Appendix D  in \cite{Peshkin}. Similarly to that case, we observe that the topological character of the Berry phase occurring in the nuclear wavefunction  produces a purely geometric gyration motion in the semiclassical expectation dynamics. In more generality, this feature is well known to apply  to the Ehrenfest equations associated to the Aharonov-Bohm problem  \cite{Peshkin,SeTa88}, although it has generated some confusion in the early stages \cite{Bocchieri,Kobe}. Given the regularized connection $\overline{\nabla\zeta}$, a generic element of its associated holonomy group \cite{Simon1983} is generated by the geometric phase
\begin{equation}\label{RGP}
\Gamma=\oint_{\gamma_0}\overline{\nabla\zeta}(\br)\cdot\de\br
\,.
\end{equation}
As anticipated above, this quantity is essentially geometric in nature as its value depends on the path $\gamma_0$. Indeed, notice that even if $\zeta$ has a pole at the location of the conical intersection, the geometric phase $\Gamma$ vanishes as the loop $\gamma_0$ encircling the singular point of $\zeta$ shrinks to the point itself. The emergence of this geometric type of  Berry phase has recently been considered in the context of quantum chemistry by Gross and collaborators \cite{Gross3,Gross2}. In some specific cases,  the authors of \cite{Gross3,Gross2} adopt an exact approach \cite{AbediEtAl2012} to the full nonadiabatic quantum molecular problem to show that the exact Berry connection quickly tends to the topological value $\oint_{\gamma_0\!}{\nabla\zeta}(\br)\cdot\de\br$ from Born-Oppenheimer theory as the loop $\gamma_0$ encircling the singular point of $\zeta$ becomes larger and larger. According to the authors of \cite{Gross2}, this feature would explain why Born-Oppenheimer theory is so successful in reproducing the values of the Berry phase observed in molecular spectroscopy experiments \cite{DeGrWhWoZw,Cocchini,vonBusch}. Whether this specific relation between $\Gamma$ and the loop $\gamma_0$ can be recovered in the present setting is an interesting question which will be approached in Section \ref{sec:JT}.

\subsection{Regularization of the diagonal correction\label{sec:RegDC}}
While the treatment in Section \ref{sec:CS} was shown to avoid singularities since the nuclear density behaves as a regularizing convolution filter, this approach crucially requires neglecting the diagonal Born-Oppenheimer correction term. As discussed in \cite{MeLe16}, the nuclear density needs to vanish at the conical intersection in order for this term to converge thereby avoiding non-integrable singularities. In turn, the appearance of this node in the wavefunction was recognized to be an artifact of the Born-Oppenheimer as this null point is eliminated in exact nonadiabatic treatments \cite{AbediEtAl2012}.

Instead of giving up including the diagonal correction in the present treatment, here we propose to model this term by applying a regularization technique that is based on the treatment in Section \ref{sec:CS}. To this end, we insist that the regularized diagonal correction term must arise from a gauge-invariant regularization of the effective electronic potential \eqref{effectivepotential}. In particular, we shall consider the density matrix formulation in the second line of \eqref{effectivepotential} in order to retain gauge-invariance at all stages. The approach from Section \ref{sec:CS} leads to the regularized electronic potential
\[
\bar\varepsilon (\phi',\nabla\phi')=\,\overline{\!E}+\frac{\hbar^2}{4M}\overline{\|\nabla\rho_\phi\|^2}
\,,
\]
where the bar denotes convolution with the nuclear density and we have used gauge-invariance to write $\bar\varepsilon (\phi',\nabla\phi')=\bar\varepsilon (\phi,\nabla\phi)$. It is clear that the non-integrable singularity carried by the second term cannot be simply regularized by a standard convolution. Thus, in order to overcome this obstacle, here we propose the replacement $\overline{\|\nabla\rho_\phi\|^2}\to \|\overline{\nabla\rho_\phi}\|^2$. This step is motivated by the fact that, if the gradient was regular, then one could invoke a sharply peaked nuclear density so that $\overline{\|\nabla\rho_\phi\|^2}\simeq \|\overline{\nabla\rho_\phi}\|^2=\|\nabla\bar{\rho}_\phi\|^2$. A similar regularization of the electronic density matrix was recently proposed also within a nonadiabatic context \cite{FoHoTr19}. In the present case, the nuclear trajectory equation reads
\begin{equation}\label{newq-eq}
M\ddot\bq+\dot\bq\times\nabla\times\overline{\nabla\zeta}(\bq)=-\nabla\bigg(\overline{\!E}(\bq)+\frac{\hbar^2}{4M}\|\overline{\nabla\rho_\phi}(\bq)\|^2\bigg).
\end{equation}
As we shall see in the context of linear vibronic coupling, this method provides a well behaved regularized version of the electronic potential, which may be modified depending on the specific profile chosen for the nuclear density.


\section{An example: the Jahn-Teller problem\label{sec:JT}}

It will be instructive to see how the ideas presented in the previous
sections work out in a concrete model: the  Jahn-Teller
Hamiltonian for linear vibronic coupling \cite{Teller37}.
Here we begin by outlining its features under the usual
Born-Oppenheimer framework  before exploring the semiclassical trajectories
approach of Section \ref{sec:CS}.

\subsection{Linear vibronic coupling}
Consider the linear $E\otimes e$ Jahn-Teller model, a simple effective
Hamiltonian which describes an electronic doublet linearly coupled to a twofold
degenerate vibrational mode; see eg. \cite{OBCh93,Yarkony}. Explicitly, the nuclear coordinates $\br=r_1\mathbf{e}_1+r_2\mathbf{e}_2+r_3\mathbf{e}_3$
are coupled to the electronic degrees of freedom through the Hamiltonian 
$
\widehat{H}=-{\hbar^{2}M^{-1}}\Delta/2
+\widehat{H}_e(\br)\,,
$
where the electronic Hamiltonian takes the form
\[
\widehat{H}_e(\br)=
\frac{k}{2}\left(r_{2}^{2}+r_{3}^{2}\right)
+g\begin{pmatrix}r_{2} & -r_{3}\\
-r_{3} & -r_{2}
\end{pmatrix}
\,,
\]
{where the last term represents the linear electronic-vibrational coupling.}
As there is cyclindrical symmetry about the $r_{1}$ axis, it will
be convenient at times to work in cylindrical coordinates 
defined by $R=\sqrt{r_{2}^{2}+r_{3}^{2}}$ and $\theta=\tan^{-1}(r_{3}/r_{2})$.
We are interested in the electronic ground state. If the nuclear coordinates
are frozen then we obtain the Hamiltonian for just the electronic
degrees of freedom, given in standard Pauli matrix notation as 
$
\widehat{H}_e={k}R^{2}/{2}+
gR\left(\cos\theta\widehat{\sigma}_3-\sin\theta\widehat{\sigma}_1\right)
$.
The energy eigenvalues for this Hamiltonian are $
E_\pm(\br)={k}R^{2}/2\pm gR
$
with a conical intersection occurring at $R=0$. Then, the ground state  wavefunction
 $|\phi(\br)\rangle$
arises from the electronic eigenvalue problem as follows:
\[
\widehat{H}_{e}
|\phi\left(\br\right)\rangle
=E(\br)
|\phi\left(\br\right)\rangle
\,,\quad\qquad\text{with}\quad\qquad
|\phi(\br)\rangle=\begin{pmatrix}\sin({\theta}/{2})\\
\cos({\theta}/{2})
\end{pmatrix},
\]
where we have denoted $E(\br):=E_-(\br)$.
Note that this ground state is double-valued since it picks up a factor of $-1$ as
$\theta$ increases from $0$ to $2\pi$. Following the Mead-Truhlar procedure,
we can construct a single-valued electronic wavefunction through the
gauge transformation $
|\phi\rangle\rightarrow|\phi\rangle'=e^{i\zeta/\hbar}|\phi\rangle$
where
$
\zeta(\br)={\hbar}\theta/2
$.
The associated magnetic potential is easily calculated, giving
\begin{equation}\label{zeta}
\nabla\zeta(\br)
=\frac{\hbar}{2R}\mathbf{e}_\theta\,,
\end{equation}
{where  $\mathbf{e}_\theta=(0,-\sin\theta,\cos\theta)^T$}.
In particular, by computing the line integral $\oint_{\gamma_R}\!\nabla\zeta\cdot\de\br$, we see that the Berry
phase associated with \emph{any} circle $\gamma_{R}$ of radius $R>0$
in the $r_{2}r_{3}-$plane (no matter how small) is simply $\pi$
- the usual topological result expected when encircling a conical
intersection. This result can also be computed from the usual delta-like representation of the associated
Berry curvature $\nabla\times \nabla\zeta=\hbar\pi\delta(r_{2})\delta\left(r_{3}\right)\mathbf{e}_1$ or, in polar coordinates,
\begin{equation}\label{curvature}
\nabla\times \nabla\zeta(\br)=\frac{\hbar}{R}\delta(R)\, \mathbf{e}_1
.
\end{equation}
Indeed, we have $\frac{1}{\hbar}\oiint_{\, \Sigma_R}\!\nabla\times \nabla\zeta\cdot\de\mathbf{S}=\pi$ for any closed surface such that $\partial\Sigma_R=\gamma_R$.

Putting these things together, we see that within the (generalized)
Born-Oppenheimer approximation we obtain a nuclear wavefunction equation
of the form \eqref{genBOeq}, 
with {$\nabla\zeta$ as given in \eqref{zeta}} and where the effective electronic
potential is
\begin{eqnarray*}
\varepsilon(\phi',\nabla\phi') = E+\frac{\hbar^2}{4M}\|\nabla\rho_\phi\|^2
= \frac{k}{2}R^{2}-gR+\frac{\hbar^{2}}{8MR^{2}}.
\end{eqnarray*}
Thus the nuclear wavefunction equation features both a singular vector
potential and a singular effective electronic potential.

\subsection{Gaussian regularization}
Now we apply the method from Section \ref{sec:CS} 
with the regularized diagonal correction as in equation \eqref{newq-eq} and by restricting to nuclear densities that are rotation-invariant, so that $
D_{0}(\boldsymbol{x})=D_0(|\boldsymbol{x}|)
$.
Then, we go on to evaluate the regularised electronic potentials involved in the method presented in Section \ref{sec:CS}. We observe that the regularised vector potential $\overline{\nabla\zeta}$ can be expressed in the form
\begin{equation}\label{RBC}
\overline{\nabla\zeta}(\bq)=g(Q)\nabla\zeta(\bq)
\,.
\end{equation}
where $Q={|\mathbf{e}_{1\!}\times\bq|}$.
To see this, note that, for any rotation matrix $\mathcal{R}$ and vector $\mathbf{a}$, 
\begin{equation*}
\overline{\mathcal{\nabla\zeta}}(\mathcal{R}\bq+\mathbf{a})=\int \!D_{0}(|\br-\mathbf{a}-\mathcal{R}\bq|)\,\nabla\zeta(\br)\,\de^{3}r=\int\! D_{0}(|\br-\bq|)\,\nabla\zeta(\mathcal{R}\br+\mathbf{a})\,\de^{3}r
\,.
\end{equation*}
{Consequently}, the regularised vector potential $\overline{\nabla\zeta}(\bq)$
inherits the cylindrical symmetry of $\nabla\zeta$ in \eqref{zeta}, i.e. symmetry
under translations in the $\mathbf{e}_1$ direction and under rotations about
$\mathbf{e}_1$. {Indeed, for any rotation $\mathcal{R}(\theta,\mathbf{e}_1)$ by $\theta$ around $\mathbf{e}_1$, we have $\nabla\zeta(\mathcal{R}(\theta,\mathbf{e}_1)\br+\kappa\mathbf{e}_1)=\mathcal{R}(\theta,\mathbf{e}_1)\nabla\zeta(\br)$ and analogously for $\overline{\mathcal{\nabla\zeta}}$.
}
So, to determine $\overline{\nabla\zeta}$ for
all $\bq$, we only need to determine $\overline{\nabla\zeta}\left(\bq\right)$ on {a reference direction orthogonal to $\mathbf{e}_1$. Upon considering} the positive $q_{3}$ axis, a direct computation {in Cartesian coordinates} shows that
\[
\overline{\nabla\zeta}(Q\mathbf{e}_3) = -\frac{\hbar}{2}\left[\int \!D_{0}(r_{1},r_{2},r_{3}-Q)\,\frac{r_{3}}{r_2^2+r_3^{2}}\,\de^{3}r\right]\mathbf{e}_{2}.
\]
On the other hand, we also have
$\nabla\zeta(Q\mathbf{e}_3)=-({\hbar}/{2})\mathbf{e}_{2}/Q$
and so we see that
$\overline{\nabla\zeta}(Q\mathbf{e}_3)=g(Q)\nabla\zeta(Q\mathbf{e}_3)$
where 
\[
g\left(Q\right)=Q\left[\int\! D_{0}(r_{1},r_{2},r_{3}-Q)\,\frac{r_{3}}{r_2^{2}+r_3^2}\,\de^{3}r\right].
\]
Then, by cylindrical symmetry, we have more generally that \eqref{RBC} holds also for $\bq$ not on the $q_3$ axis.

{Therefore, equation \eqref{RBC} identifies} the regularized vector potential as a $Q$-dependent rescaling of the original with the scaling factor given by $g(Q)$.
{We observe  that, for a circle of radius
$Q$ in the $q_{2}q_{3}-$plane, the Berry phase \eqref{RGP} obtained from the regularized Berry connection \eqref{RBC}  is given by 
\begin{equation}\label{phase}
\Gamma=\pi g(Q)
\end{equation}
rather than the original $\pi$.}
In turn, equation \eqref{RBC} implies that the regularized Berry curvature has the form
\begin{equation}\label{RBCurv}
\nabla\times\overline{\nabla\zeta}(\bq)=\frac\hbar2\frac{g'(Q)}{Q}\mathbf{e}_1
\,,
\end{equation}
so that the equation \eqref{newq-eq} for nuclear motion becomes
\begin{equation}\label{newq-eq2}
M\ddot\bq=\frac\hbar2\frac{g'(Q)}{Q}\,\mathbf{e}_1\times\dot\bq-\nabla\bigg(\overline{\!E}(\bq)+\frac{\hbar^2}{4M}\|\overline{\nabla\rho_\phi}(\bq)\|^2\bigg).
\end{equation}

In order to understand the physical content of this equation, we proceed by specializing the nuclear density to a Gaussian profile
\begin{equation}\label{Gaussianprofile}
D_0(\boldsymbol{x})=\frac{1}{\big(\sqrt{\pi\hbar}\lambda\big)^{3}}\exp\left(-\frac1\hbar\frac{|\boldsymbol{x}|^{2}}{\lambda^{2}}\right)
,
\end{equation}
where $\sqrt{\hbar}\lambda$ is a lengthscale to be fixed depending on modeling purposes.
In this case, we have
\begin{equation}\label{BCon}
g(Q)=1-\exp\!\left(-\frac1\hbar\frac{Q^2}{\lambda^2}\right)
\ \implies\
\frac\hbar2\frac{g'(Q)}{Q}=\frac{1}{\lambda^2}\exp\!\left(-\frac1\hbar\frac{Q^{2}}{\lambda^{2}}\right)
\end{equation}
{and thus \eqref{phase} gives
\begin{equation}\label{phase-bis}
\Gamma=\pi\!\left[1-\exp\!\left(-\frac1\hbar\frac{Q^{2}}{\lambda^{2}}\right)\right],
\end{equation}
so that the geometric phase rapidly reaches the topological index $\pi$ as $Q$ increases and the loop expands away from the original singularity.}
In turn, the regularized Berry curvature  \eqref{RBCurv} recovers the delta-like singular case in the limit $\lambda\to0$.

In addition, after some calculations using modified Bessel functions, we obtain
\begin{equation}\label{PESbar}
\overline{\!E}(\bq)=\hbar{\lambda^2}\frac{k}2 \left(1+\frac{Q^2}{\hbar\lambda^2}\right)-\sqrt{\pi\hbar}\lambda \,\frac{g}{2}\left[\left(1+\frac{Q^2}{\hbar\lambda^2}\right)I_0\!\left(\frac{Q^2}{2\hbar\lambda^2}\right)+\frac{Q^2}{\hbar\lambda^2}I_1\!\left(\frac{Q^2}{2\hbar\lambda^2}\right)\right]{e^{-\frac{Q^2}{2\hbar\lambda^2}}}
.
\end{equation}
By proceeding analogously, we also have
\begin{equation}\label{diagbar}
\frac{\hbar^2}{4M}\|\overline{\nabla\rho_\phi}(\bq)\|^2=\frac{\pi\hbar}{16M\lambda^2}\left[I_0^2\!\left(\frac{Q^2}{2\hbar\lambda^2}\right)+I_1^2\!\left(\frac{Q^2}{2\hbar\lambda^2}\right)\right]{e^{-\frac{Q^2}{\hbar\lambda^2}}},
\end{equation}
which completes the evaluation of the potential force term in \eqref{newq-eq2} upon recalling $Q=|\mathbf{e}_1\times\bq|$.

\subsection{Comparison with exact nonadiabatic studies}
The expressions in  \eqref{BCon}, \eqref{PESbar} and \eqref{diagbar} represent the regularised potential energy surface (including the diagonal Born-Oppenheimer correction term) and vector potential computed in the special case of a Gaussian nuclear density profile. We have managed to derive explicit expressions for these quantities from which it is clear that they are indeed well-behaved at the conical intersection $Q=0$. For example, the double-cone structure of the electronic energy surface $E_\pm(\bq)$ is smoothened and the diagonal Born-Oppenheimer correction is rendered finite.

\begin{figure}[h]
\center
\includegraphics[width=0.5\textwidth]{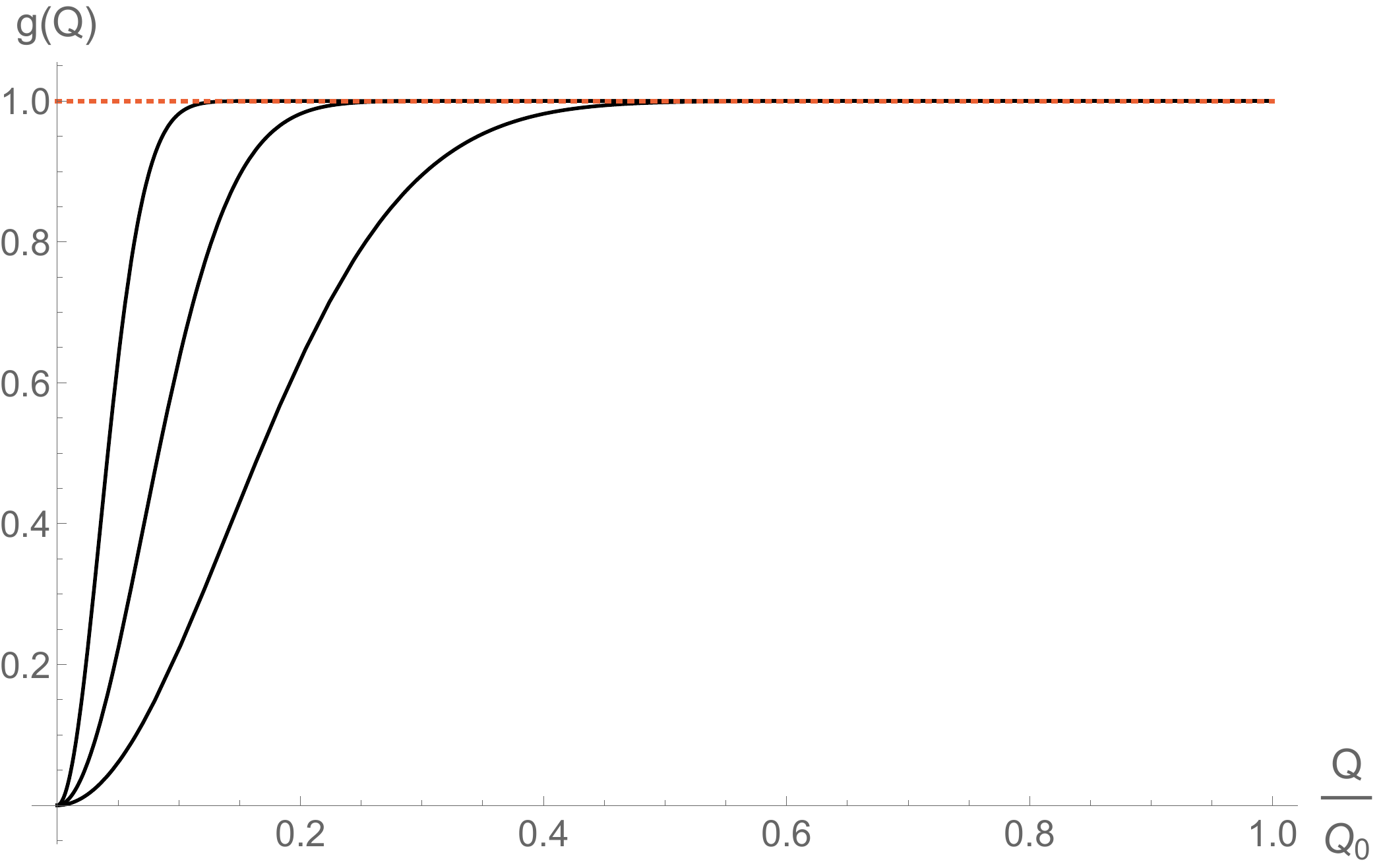}
\caption{\footnotesize Plot of $g(Q)$ for $\varepsilon=1/5,1/10,1/20$ (black, solid), compared with the BO limit (red, dotted).}
\label{gfunctions}
\end{figure}

A particularly interesting aspect of the regularised vector potential is the behavior of the associated Berry phase \eqref{phase} upon encircling the conical intersection along a circular loop. As the loop gets bigger, the phase rapidly tends to the topological result of $\pi$, since the proportionality factor $g(Q)$ between the regularised and usual Berry connection rapidly tends to 1 as $Q\rightarrow\infty$ as is clear from \eqref{BCon}. This situation is illustrated in Fig. \ref{gfunctions}, which shows the behavior of the scaling factor $g(Q)=1-\exp(-\epsilon^{-2}{Q^2}/{ Q_0^2})$. In analogy to the study in \cite{Gross3}, here we have introduced  $\epsilon^2={\hbar^2M^{-1} k^{3}}/{g^4}$ and $Q_0=g/k$. While $\epsilon$ quantifies the degree of nuclear density localization, $Q_0$ is  the radius at which the original energy surface $E$ reaches its minimum. The dependence of the phase on the radius of the circle reflects the fact that this quantity is essentially geometric in nature. 
As discussed in Section \ref{sec:BP}, the emergence of a similar path-dependent geometric phase in quantum chemistry has already been established by Gross and collaborators \cite{Gross3,Gross2} in the context of the exact wavefunction factorization for nonadiabatic dynamics \cite{AbediEtAl2012}. In fact, in  \cite{Gross3} the exact Berry phase was computed for the case of the linear $E\otimes e$ Jahn-Teller model and found to increase from $0$ to $\pi$ as $Q\rightarrow\infty$. In this setting, the lengthscale \eqref{qw} identifies  the characteristic width of the peak in the Berry curvature. The authors concluded that this behavior, having been derived within the nonadiabatic framework of the exact factorization, is a \emph{non-adiabatic effect}. Intriguingly, our regularised Berry phase displays exactly the same characteristic behavior, but we have recovered the result within an \emph{adiabatic} treatment. In fact, for us the Berry phase tends to the asymptotic topological value of $\pi$ extremely rapidly owing to the Gaussian nuclear density profile, representing a faster approach to the limit than in the exact factorization approach and maintaining consistency with experimental measurements which are, so far, not sensitive to the deviations from $\pi$. 

Of course, the precise rate at which the Berry phase approaches $\pi$ depends on our choice of nuclear density profile. For example, besides adopting a Gaussian regularisation,  we could consider the (infinite) family of generalized normal distributions which have the functional form
$
D_0^{(\beta)}(\boldsymbol{x})={\beta}\big[{4\pi\Gamma({3}/{\beta})(\sqrt{\hbar}\lambda)^{3}}\big]^{-1\!}\exp\!\big[-({\lambda^{-1}|\boldsymbol{x}|}/{\sqrt{\hbar}})^{\beta}\big]
$
\noindent
depending on a shape parameter $\beta$ and written in terms of Euler's $\Gamma-$function. The Gaussian regularization \eqref{Gaussianprofile} simply corresponds to the choice $\beta=2$, while the case $\beta=1$ identifies the cusp-like Laplace distribution.
For the general case, the regularised Berry connection is a rescaled version of the original with the scaling factor $g(Q)$ introduced in \eqref{RBC}.
%
We note that the Laplace distribution falls off less rapidly (though still exponentially) compared to the Gaussian, and this leads to a Berry phase which tends more slowly to the asymptotic topological phase of $\pi$.

\begin{figure}[h]
\centering
\includegraphics[width=0.5\textwidth]{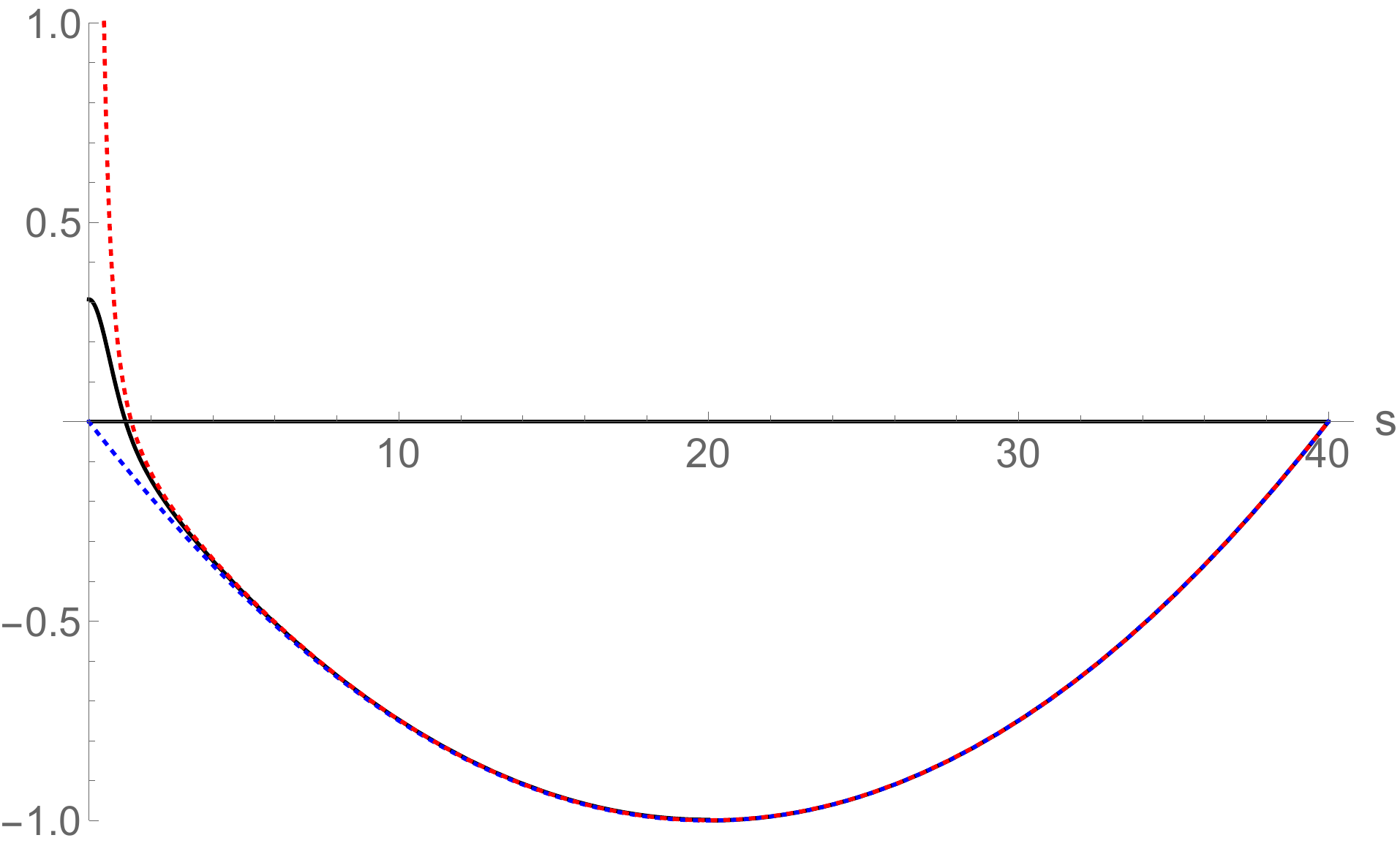}
\caption{\footnotesize Plot of effective potential $\Delta^{-1}({\,\overline{\!E}}+{\hbar^2M^{-1}}\|\overline{\nabla\rho_\phi}\|^2/4)$ for $\varepsilon=1/20$ (black, solid), compared with the BO potential with (red, dotted, upper) and without (blue, dotted, lower) the diagonal correction.}
\label{diagon}
\end{figure}

Another point of discussion is given by the  diagonal correction regularization introduced in Section \ref{sec:RegDC}. Following the analysis in \cite{Gross3}, we now pick the width of the nuclear wavepacket to be given by 
\begin{equation}
\hbar\lambda^2=\frac{\hbar^2 k}{g^2M}
\,,
\label{qw}
\end{equation}
whose meaning will be clarified later. 
Then, the relations  \eqref{BCon}, \eqref{PESbar} and \eqref{diagbar} are expressed in terms of dimensionless variable $s={Q}/{\sqrt{\hbar \lambda^2}} $, the Jahn-Teller {\it stabilization energy}  $\Delta=k^{-1}{g^2}/{2}$, and the previously introduced ratio $\epsilon$. For example, \eqref{PESbar} gives
\[
{\,\overline{\!E}}/\Delta=\epsilon^2 {(1+s^2)}-\sqrt{\pi}\epsilon \left[(1+s^2)I_0({s^2}/{2})+s^2 I_1({s^2}/{2})\right]e^{-{s^2}/{2}}
,
\]
while \eqref{diagbar} gives an analogous expression of the diagonal correction ${\hbar^2M^{-1}\Delta^{-1}}\|\overline{\nabla\rho_\phi}\|^2/4$, which appears to be independent of $\epsilon$. Then, the regularized effective electronic potential can be compared directly to its Born-Oppenheimer correspondent; see  figure \ref{diagon}.
\rem{ 

\begin{equation}
\frac{\hbar^2}{4M\Delta}\|\overline{\nabla\rho_\phi}(\bq)\|^2=\frac{\pi}{8}\left[I_0^2\!\left(\frac{s^2}{2}\right)+I_1^2\!\left(\frac{s^2}{2}\right)\right]{e^{-s^2}}
\end{equation}

\begin{equation}
\frac{\overline{\!E}_-(\bq)}{\Delta}=\varepsilon^2 \left(1+s^2\right)-\sqrt{\pi}\varepsilon \,\left[\left(1+s^2\right)I_0\!\left(\frac{s^2}{2}\right)+s^2 I_1\!\left(\frac{s^2}{2}\right)\right]{e^{-\frac{s^2}{2}}}
\end{equation}

\begin{equation}
g(Q)=1-\exp\!\left(-\frac{Q^2}{\varepsilon^2 Q_0^2}\right)=1-\exp\!\left(-s^2\right).
\end{equation}

\noindent
in terms of the dimensionless variable $s=\frac{Q}{Q_{\mathrm{width}}}$ and where we have introduced the dimensionless ratio $\varepsilon^2=\frac{\hbar^2 k^{3}}{g^4 M}$ as well as the Jahn-Teller stabilization energy  $\Delta=\frac{g^2}{2k}$ and the length scale $Q_0=g/k$. 
} 
We observe that the effective potential obtained from the regularization in Section \ref{sec:RegDC} is again strikingly similar to the exact energy surface recently presented in \cite{Gross3} and the two profiles exactly coincide away from the original singularity. In Fig.~7 of \cite{Gross3}, the exact energy surface appears to reach the vertical axis at slightly higher energy values, within the interval $(0.5,0.75)$. This result indicates that the proposed adiabatic model reproduces all the essential exact nonadiabatic results quite faithfully.

Notice that our study has focused on linear vibronic coupling, while higher-order vibronic interactions may become important in more realistic situations thereby  warping the adiabatic energy surfaces whose rotational symmetry is then broken. In this case, the conical intersection at the origin  is accompanied by additional singular points whose contributions have been studied in e.g. \cite{ZwFr87} for the occurrence of both linear and quadratic coupling. The application of the present model to these more involved scenarios is left for future work. Here, we shall simply point out that retaining quadratic coupling terms in the $E\otimes e$ Jahn-Teller problem leads to no essential difficulties, since the Berry connection has locally the same profile nearby any of the four conical intersections.

\section{Conclusions}

This paper has presented a new Born-Oppenheimer molecular dynamics approach in the presence of conical intersections. While in this context the usual quantum-classical picture leads to intractable nuclear trajectory equations, here we have replaced the classical limit with a closure of the nuclear Ehrenfest equations that is based on the variational formulation of  Bohmian mechanics. Instead of following well-established approaches based on coherent states \cite{KramerSaraceno,Shi,ShiRabitz}, here we have focused on nuclear Bohmian trajectories which were conveniently restricted to simple translations of the nuclear configuration coordinates. This Bohmian closure leads to a nontrivial regularized Berry connection given by the convolution of its original singular expression with the reference nuclear density. The corresponding regularized  curvature appears as a Lorentz-force term in the nuclear trajectory equations, which then accounts for holonomy effects. In this context, the regularization process leads to a path-dependent geometric phase replacing the usual topological index, thereby recovering the same phenomenology as in recent  nonadiabatic studies \cite{Gross2,Gross3}.

We showed that the Bohmian closure alone is not sufficient to deal with the well-known problems arising from the diagonal Born-Oppenheimer correction, which leads to divergent integrals. In order to tackle situations where the diagonal correction is not neglected, we proposed a further regularization operating on the gauge-invariant form of the effective electronic potential \eqref{effectivepotential}. This further regularization step consists in the replacement $\overline{\|\nabla\rho_\phi\|^2}\to\|\overline{\nabla\rho_\phi}\|^2$ and  is motivated by the regular case in which  $\rho_\phi$ is differentiable and the reference nuclear density is sufficiently peaked.

In order to illustrate the essential features of the proposed scheme, we have applied our model to the Jahn-Teller problem for linear vibronic coupling. In this case, we showed that a rotation-invariant nuclear density leads to a regularized Berry connection that is a rescaled variant of the original singular expression by a smooth function of the radial coordinate in cylindrical symmetry. Then, we showed that for a generalized class of normal density distributions the Berry phase rapidly reaches the topological index value as the loop encircling the singularity gets bigger. In general, the slope with which the topological index is reached is slower as the nuclear reference distribution gets wider. This qualitative  behavior is precisely the same as that recently found in exact nonadiabatic studies \cite{Gross3}, thereby indicating how the proposed model is successful in reproducing the same phenomenology as in the nonadiabatic fully quantum case.

Within the context of the $E\otimes e$ Jahn-Teller problem, we also illustrated the behavior of the regularized effective energy surface. In particular, by a convenient nondimensionalization, we presented again a substantial similarity between the regularized energy and the exact energy surface obtained from a fully quantum nonadiabatic treatment \cite{Gross3}. Once more, this result supports the validity of our model even in the situations when the diagonal correction term is retained, thereby eliminating the highly pathological behavior otherwise appearing in the original treatment. 

While the application of our molecular dynamics scheme is left for future studies, here we point out that the inclusion of quadratic vibronic coupling in the Jahn-Teller problem is not expected to lead to  essential difficulties. This is due to the fact that the additional conical intersections emerging in that case \cite{ZwFr87} produce a Berry connection that is locally the same in the neighborhood of the singular points.  We also note that the Berry connection is easily computed in the Jahn-Teller model by exploiting the high degree of symmetry, while the treatment of realistic polyatomic systems will require a more systematic approach. For example, following the technique proposed in \cite{MaZhGuYa16,XiMaYaGu17}, one could approximate the Berry connection through performing a fitting to a two-state diabatic representation. Once the Berry connection is calculated, it should be possible to apply our regularization to complicated polyatomic systems just as in the Jahn-Teller model.

Before closing, we emphasize that, despite the striking similarities with the results in \cite{Gross3}, the exact nonadiabatic factorization therein is conceptually very different from the model proposed here. Indeed, while in \cite{Gross3} the electronic energy surface  appears to be smooth as a result of the exact nonadiabatic approach, in our case the treatment is essentially adiabatic so that the smoothing process arises as a modeling strategy motivated by the nuclear Ehrenfest dynamics. Thus, we expect the  model proposed here to retain  limitations from the standard adiabatic treatment underlying the present study. For example, it has been shown \cite{Gross1} that there are cases in which the nonadiabatic factorization leads to an exactly  vanishing Berry phase due to the appearance of an electronic wavefunction that is both real and single-valued. These and other features escape from the present treatment, which however was shown to retain some of the nonadiabatic effects eliminating topological singularities from the nuclear problem.

\paragraph{Acknowledgements.} We are grateful to Denys Bondar, Basile Curchod, Michael Foskett, Darryl Holm, Artur Izmaylov, and Ryan Requist for stimulating discussions and correspondence. The work of JIR was supported by the EPSRC grant EP/P021123/1, while CT acknowledges partial support from the LMS grant 41909.

\end{document}